\documentclass[aps,amsmath,amssymb,superscriptaddress,prb,10pt,twocolumn,eqrefnum]{revtex4-2}
\usepackage{amsmath}    
\usepackage{graphicx}  
\usepackage{verbatim}  
\usepackage{color}      
\usepackage{subfigure} 
\usepackage{hyperref}   
\usepackage{xcolor,txfonts,mathtools,mathrsfs}
\usepackage{bm}   
\usepackage{soul}
\usepackage{float}

\definecolor{SC-color}{named}{blue}

\definecolor{SCh-color}{named}{green}

\newcommand{\beq}{\begin{equation}}
\newcommand{\eeq}{\end{equation}}
\newcommand{\Beq}{\begin{eqnarray}}
\newcommand{\Eeq}{\end{eqnarray}}
\newcommand{\bml}{\begin{multline}}

\newcommand{\bea}{\begin{align}}
\newcommand{\ena}{\end{align}}
\newcommand{\bsp}{\begin{split}}
\newcommand{\esp}{\end{split}}

\begin{document}
\title{DC Josephson effect between two Yu-Shiba-Rusinov bound states}
\author{Subrata Chakraborty}
\email[Correspondence to: ]{subrata.chakraborty@uni-konstanz.de}
\affiliation{Fachbereich Physik, Universit{\"a}t Konstanz, D-78457 Konstanz, Germany}
\author{Danilo Nikoli\'c}
\email{danilo.nikolic@uni-konstanz.de}
\affiliation{Fachbereich Physik, Universit{\"a}t Konstanz, D-78457 Konstanz, Germany}
\affiliation{Institut f\"ur Physik, Universit\"at Greifswald, Felix-Hausdorff-Strasse 6, 17489 Greifswald, Germany}
\author{Rub\'en Seoane Souto}
\affiliation{Departamento de F\'{\i}sica Te\'orica de la Materia Condensada and Condensed Matter Physics Center (IFIMAC),
Universidad Aut\'onoma de Madrid, E-28049 Madrid, Spain}
\affiliation{Instituto de Ciencia de Materiales de Madrid (ICMM), Consejo Superior de Investigaciones Cient\'{\i}ficas (CSIC),
Sor Juana In\'es de la Cruz 3, 28049 Madrid, Spain}
\author{Wolfgang Belzig}
\affiliation{Fachbereich Physik, Universit{\"a}t Konstanz, D-78457 Konstanz, Germany}
\author{Juan Carlos Cuevas}
\affiliation{Departamento de F\'{\i}sica Te\'orica de la Materia Condensada and Condensed Matter Physics Center (IFIMAC),
Universidad Aut\'onoma de Madrid, E-28049 Madrid, Spain}

\date{\today}

\begin{abstract} 
Motivated by recent experiments [Nat. Phys. {\bf 16}, 1227 (2020)], we present here a theoretical study of the DC Josephson 
effect in a system comprising two magnetic impurities coupled to their respective superconducting electrodes and which exhibit
Yu-Shiba-Rusinov (YSR) states. We make use of a mean-field Anderson model with broken spin symmetry to compute the supercurrent in 
this system for an arbitrary range of parameters (coupling between the impurities, orientation of the impurity spins, etc.). We predict 
a variety of physical phenomena such as (i) the occurrence of multiple $0$-$\pi$ transitions in the regime of weak coupling that 
can be induced by changing the energy of the YSR states or the temperature; (ii) the critical current strongly depends on the relative
orientation of the impurity spins and it is maximized when the spins are either parallel or antiparallel, depending on the 
ground state of the impurities; and (iii) upon increasing the coupling between impurities, triplet superconductivity is generated 
in the system and it is manifested in a highly nonsinusoidal current-phase relation. In principle, these predictions can be 
tested experimentally with the existing realization of this system and the main lessons of this work are of great relevance for 
the field of superconducting spintronics. 
\end{abstract}

\maketitle

{\section{Introduction}} The advent of scanning tunneling microscopy (STM) has enabled the investigation of the interplay between 
magnetism and superconductivity at the atomic scale in the context of single magnetic impurities on superconducting surfaces 
\cite{Heinrich2018,Choi2019}. One of the most emblematic manifestations of this interplay is the appearance of in-gap superconducting 
bound states, known as Yu-Shiba-Rusinov (YSR) states \cite{Yu1965,Shiba1968,Rusinov1969}. These bound states can induce a quantum 
phase transition in which the ground state of the impurity system may change between a singlet state (spin $0$) and a 
double state (spin $1/2$) \cite{Balatsky2006,Franke2011,Bauer2013,Malavolti2018,Farinacci2018}. The change in fermion parity 
in this transition is accompanied by a supercurrent reversal in Josephson junctions, often referred to as $0$-$\pi$ transition. This 
kind of transition has been reported in different mesoscopic systems ranging from superconductor-ferromagnet hybrid junctions 
\cite{Ryazanov2001,Bauer_PRL2004,Robinson2006,Kontos2002,Caruso_PRL2019,Razmadze_PRB2023} to single~\cite{vanDam2006,
DeFranceschi2010,Martin-Rodero2011,Meden2019,Razmadze_PRL2020,Bargerbos_PRXq2022} and double~\cite{Su_NatComm2017,Grove-Rasmussen2018,
Estrada_PRL2018,Estrada2020,Bouman_PRB2020,Vekris_PRR2021,Kurtossy_NanoLet2021,Steffensen_PRB2022,Zhang_PRL2022,Debbarma_PRB2022} 
quantum dots coupled to superconducting leads. However, in the context of STM-based experiments with magnetic impurities on
superconductor substrates, it has been very elusive to observe a $0$-$\pi$ transition due to the lack of phase sensitivity in this 
type of setup. This was recently circumvented with the help of the presence of a second channel in an experiment featuring a single 
YSR pair of states in a vanadium-based system \cite{Karan2022}. 
\begin{figure}[h]
\includegraphics[width=1\linewidth]{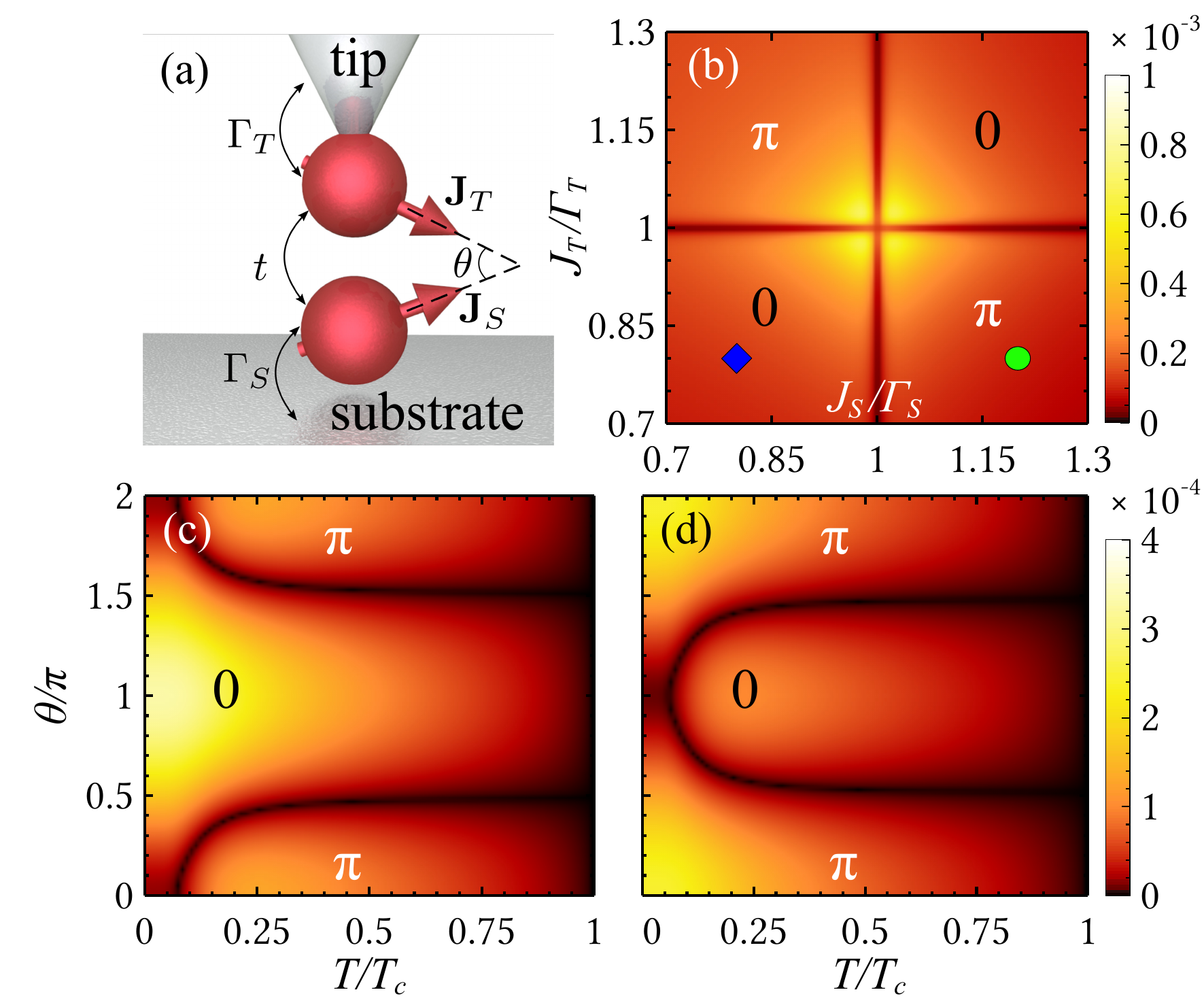}
\caption{  (a) Schematic representation of the system under study. Two magnetic impurities with
exchange fields $\boldsymbol{J}_S$ and $\boldsymbol{J}_T$ forming an angle $\theta$ are respectively coupled to a 
superconducting substrate and to a superconducting STM tip. The tunneling rates $\Gamma_T$ and $\Gamma_S$ measure 
the strength of the coupling of the impurities to the tip and substrate, respectively, and $t$ is the hopping matrix 
element describing the tunnel coupling between the impurities. (b) The zero-temperature critical current as a function of
the ratios $J_S/\Gamma_S$ and $J_T/\Gamma_T$ for $\theta=\pi/2$. In panel (b), we mark  ($J_S/\Gamma_S=0.8$, $J_T/\Gamma_T=0.8$) 
(blue colored diamond shaped) and ($J_S/\Gamma_S=1.2$, $J_T/\Gamma_T=0.8$) (green colored circular shaped) points. (c) The critical 
current as a function of the temperature and the angle $\theta$ for $J_S/\Gamma_S = J_T/\Gamma_T = 0.8$. (d) The same quantity as 
in panel (c), but for $J_S/\Gamma_S=1.2$ and $J_T/\Gamma_T=0.8$. For panels (b)-(d) $t = \Delta_0$, $\Gamma_{S/T} = 100\Delta_0$, 
$U_{S/T} = 0$ and $\gamma=0.003\Delta_0$. The critical currents with colormaps in panels (b)-(d) are expressed in the units of 
$e\Delta_0/\hbar$.}
\label{fig-scheme}
\end{figure} 

In this work we focus on a system in which two magnetic impurities are coupled to their respective superconducting 
electrodes such that one can have quantum tunneling between two individual YSR states, see Fig.~\ref{fig-scheme}(a). This 
system has been recently realized experimentally in the context of STM and it constitutes the ultimate limit of quantum 
tunneling \cite{Huang2020a}. In those experiments the authors focused on the current-voltage characteristics and the 
results have been understood with the help of mean-field models that account for the relative orientation between the impurity 
spins \cite{Villas2021,Huang2021,Ohnmacht2023}. Here, we shall focus on the DC Josephson effect and employ the mean-field model 
to make concrete predictions for the current-phase relation and the critical current in this two-impurity system. Our study is 
motivated by the fact that this system is an ideal playground to investigate fundamental questions in the field of superconducting 
spintronics \cite{Linder2015,Eschrig2015}. In particular, we predict that the supercurrent in this hybrid atomic-scale system 
exhibits a number of spin-related phenomena among which we can highlight: (i) the occurrence of multiple $0$-$\pi$ transitions 
in the tunneling regime (weak coupling between the impurities) that can be induced by changing the position of the YSR states or 
by modifying the temperature; (ii) the critical current depends drastically on the relative orientation of the impurity spins
and it is maximized either when the impurity spins are parallel or antiparallel, depending on the ground state of the impurities; 
(iii) upon increasing the coupling between the impurities, triplet superconductivity is generated and is revealed in a 
highly nonsinusoidal current-phase relation. These predictions could be tested experimentally using the exact system of 
Ref.~\cite{Huang2020a} and our main conclusions have important implications for the whole field of superconducting spintronics.

The rest of the paper is organized as follows. First, in Sec.~\ref{sec-model} we present our theoretical model 
to describe the coupling between two YSR states, and provide a prescription to compute the DC Josephson current between the two 
states. Then, in Sec.~\ref{sec-results} we discuss our main results on the $0$-$\pi$ transitions in the supercurrent for low 
transmissions, the emergence of triplet superconductivity and how it is reflected in the supercurrent, and the appearance of 
higher order harmonics in the current-phase relation beyond the tunnel regime. We summarize the main conclusions of 
this work in Sec.~\ref{sec-conclusions}. Some of the technical details related to the supercurrent in the tunneling regime are
reported in Appendix~\ref{AppenA}, while a discussion of the critical current in the case of nonorthogonal spin 
orientations is presented in Appendix~\ref{AppenB}.

\vspace{2mm}

\section{Model} \label{sec-model}

\subsection{System}
To study the supercurrent between two YSR states, we consider the two-impurity system schematically represented in 
Fig.~\ref{fig-scheme}(a). Here, a magnetic impurity is attached to a superconducting substrate ($S$) and another impurity 
decorates a superconducting STM tip ($T$). The system was already realized in Refs.~\cite{Huang2020a,Huang2021}. Because of the 
strong coupling to their respective superconducting electrodes, both impurities exhibit a pair of YSR states. In turn, both 
impurities are tunnel coupled, which causes YSR-hybridization and the corresponding supercurrent flow when a phase difference 
is established across the junction. To describe this system,  we make use of the mean-field model used in Ref.~\cite{Villas2021}, 
which is summarized by a Hamiltonian given by $\hat{{\bar{H}}} = \hat{{\bar{H}}}_S + \hat{{\bar{H}}}_T + \hat{{\bar{V}}}$. Here,
$\hat{{\bar{H}}}_{S,T}$ describes the subsystem formed by an impurity attached to the superconductor $j = S,T$ and 
$\hat{{\bar{V}}}$ describes the coupling between the magnetic impurities. With respect to the global spin-quantization 
axis, defined as the middle angle between ${\bf{J}}_T$ and ${\bf{J}}_S$, we can choose the basis set for an impurity attached 
to the respective superconductor $j = S,T$ given by ${\bar{d}}^\dag_j = ({\bar{d}}^\dag_{j\uparrow}, {\bar{d}}_{j\downarrow},
{\bar{d}}^\dag_{j\downarrow}, -{\bar{d}}_{j\uparrow})$. For the superconducting electrodes we choose a basis set with respect 
to the global spin-quantization frame as ${\bar{c}}^\dag_{{\bf{k}}j} = ({\bar{c}}^\dag_{{\bf{k}}j\uparrow}, 
{\bar{c}}_{{\bf{k}}j\downarrow}, {\bar{c}}^\dag_{{\bf{k}}j\downarrow}, -{\bar{c}}_{{\bf{k}}j\uparrow})$. The Hamiltonian of 
a bare YSR system, i.e., a magnetic impurity plus the corresponding superconductor is expressed in the above basis as 
$\hat{{\bar{H}}}_j = \hat{{\bar{H}}}_{{\rm{elec}},j} + \hat{{\bar{H}}}_{{\rm{imp}},j} + \hat{{\bar{H}}}_{{\rm{int}},j}$, where
\begin{eqnarray}
&& \hat{{\bar{H}}}_{{\mathrm{imp}},j} = \frac{1}{2} {\bar{d}}^\dag_j ~{\bar{H}}_{{\rm{imp}},j}~{\bar{d}}_j, \label{eq2} \\
&& \hat{{\bar{H}}}_{{\rm{elec}},j} = \frac{1}{2}\sum_{\bf{k}}~{\bar{c}}^\dag_{{\bf{k}}j}~{{\bar{H}}}_{{\rm{elec}},{\bf{k}}j}~ 
{\bar{c}}_{{\bf{k}}j},  \label{eq3} \\
&& \hat{{\bar{H}}}_{{\rm{int}},j} = \frac{1}{2}\sum_{{\bf{k}}} {\bar{c}}^\dag_{{\bf{k}}j}~{{\bar{H}}}_{{\rm{int}},j}~{\bar{d}}_j + 
{\rm{H.c.}}. \label{eq4}
\end{eqnarray}
In Eqs.~\eqref{eq2}-\eqref{eq4} the Hamiltonian matrices are given by ${\bar{H}}_{{\rm{imp}},j} = U_j(\sigma_0  \tau_3) + 
{\bf{J}}_j \cdot ({\bm{\sigma}} \tau_0)$, ${{\bar{H}}}_{{\rm{elec}},{\bf{k}}j}= \sigma_0  (\xi_{{\bf{k}}j} \tau_3 + 
\Delta_j e^{i\phi_j \tau_3}\tau_1)$, and ${{\bar{H}}}_{\mathrm{int},j} = v_j (\sigma_0  \tau_3)$. They define the bare 
magnetic impurity attached to the respective superconductor $j$, the corresponding superconducting electrode $j$, and 
the interaction between them, respectively. These Hamiltonian matrices are expressed in spin $\otimes$ Nambu space $\sigma_i\tau_j$, where $\sigma_i$ and $\tau_j$ denote the Pauli matrices in the respective space.
The superconductors are described by electronic energy $\xi_{{\bf{k}}j}$, pairing potential $\Delta_j$, and superconducting 
phase $\phi_j$. The bare magnetic impurities are described by the single-particle energy $U_j$ and the exchange field 
${\bf{J}}_j$. An important parameter of this model is the angle $\theta$ describing the relative orientation between the 
two impurity spins, i.e., between ${\bf{J}}_S$ and ${\bf{J}}_T$. The strength of the coupling between an impurity and its
corresponding superconductor is denoted by $v_j$. Due to the hybridization between the two YSR systems the tunneling Hamiltonian 
in the above basis adopts the form
\Beq
\hat{{\bar{V}}} = \frac{1}{2} {\bar{d}}^\dag_T \bar{V}_{TS} {\bar{d}}_S +\frac{1}{2} {\bar{d}}^\dag_S \bar{V}_{ST} 
{\bar{d}}_T, \label{eq8}
\Eeq
where ${\bar{V}}_{TS}=t(\sigma_0 \tau_3)= \bar{V}_{ST}$ and $t$ is the hopping matrix element that describes the strength 
tunneling between the two YSR subsystems. The tunneling Hamiltonian $\hat{{\bar{V}}}$ accounts for the spin-independent 
tunneling processes. However, spin-flip process will effectively take place when the spins are misaligned, as we explain in 
what follows.

Instead of working with a global spin-quantization frame, for convenience, we use mixed quantization axes such that $S$ and 
$T$ subsystems are spin-quantized along the exchange fields ${\bf{J}}_S$ and ${\bf{J}}_T$, respectively. For this
purpose we make use of the rotation matrix for subsystem $j = S,T$ as $R_j = \exp\left({i \theta_j \sigma_2/2}\right) \tau_0$, 
where $\theta_{T/S}=\pm\theta/2$ is the relative angle of the exchange fields ${\bf{J}}_{T/S}$ with respect to the global 
quantization axis. Upon rotation, we can define the basis along the respective exchange field directions as 
${{d}}^\dag_j = R_j {\bar{d}}^\dag_j \equiv ({{d}}^\dag_{j\uparrow}, {{d}}_{j\downarrow}, {{d}}^\dag_{j\downarrow}, 
-{{d}}_{j\uparrow})$. Due to the superconductor-impurity coupling, the dressed retarded/advanced ($r/a$) Green's function 
matrix of each impurity in the basis $d^\dag_j$ becomes ${g}^{r/a}_{jj}(E) = {g}^{r/a}_{jj\uparrow \uparrow}(E) \oplus
{g}^{r/a}_{jj\downarrow \downarrow}(E)$ with
\begin{widetext}
\Beq
&& {g}^{r/a}_{jj \sigma\sigma}(E) = \frac{1}{D_{j\sigma}(E)} \left(\begin{array}{cccc}
    E\Gamma_j +(E +U_j -J_{j\sigma})\sqrt{\Delta^2_j -E^2} & \Gamma_j\Delta_je^{i\phi_j} \\
     \Gamma_j\Delta_je^{-i\phi_j}   & E\Gamma_j +(E -U_j -J_{j\sigma})\sqrt{\Delta^2_j -E^2}
    \end{array}\right), \label{eq10}
\Eeq
\end{widetext}
where $\Gamma_j=\pi N_{0,j}v^2_j$ ($N_{0,j}$ being the normal density of states of electrode $j$), 
$D_{j\sigma}(E) = 2\Gamma_jE(E-J_{j\sigma}) + \left[(E-J_{j\sigma})^2-U^2_j - \Gamma^2_j\right] 
\sqrt{\Delta^2_j -E^2}$, and $E=E\pm i\gamma$ (with $\gamma\rightarrow 0^+$). Here, $J_{j\uparrow}= 
+J_j$ and $J_{j\downarrow}=-J_j$. The YSR bound states can be obtained by setting $D_{j\sigma}(E) = 0$ for $J_j,
\Gamma_j\gg\Delta_j$, and they are expressed as $E_{{\rm YSR},j\uparrow}=-E_{{\rm YSR},j\downarrow}$ with
\beq
E_{{\rm YSR},j\uparrow} = \Delta_j \frac{J^2_j-\Gamma^2_j-U^2_j}{\sqrt{\left[\Gamma^2_j +(J_j-U_j)^2\right]
\left[\Gamma^2_j + (J_j+U_j)^2\right]}}. \label{eq11}
\eeq
For $U_j=0$ it is easy to show that if $J_j/\Gamma_j\lessgtr1$, then $E_{{\rm YSR},j\uparrow} = -E_{{\rm YSR},j\downarrow} 
\lessgtr0$. Next, to study the supercurrent between the two impurities we need the tunneling matrices with respect to the 
mixed quantization frames. The tunneling matrices are transformed as $V_{TS} = R_T~{\bar{V}}_{TS}~ R^\dag_S$ and $ V_{ST} = 
R_S~{\bar{V}}_{ST}~ R^\dag_T$. With this transformation, the coupling matrices acquire nondiagonal elements in spin space, 
which means that effectively we have a spin-active interface in which there are spin-flip processes whose probabilities 
depend on the relative orientation of the impurity spins described by the angle $\theta$.

{\subsection{Supercurrent}}
Our goal is to understand the supercurrent in our two-impurity system. For this purpose, we employ standard nonequilibrium 
Green's function techniques, see, e.g., Ref.~\cite{Villas2021}, and express the zero-bias current in terms of Keldysh-Green's 
functions as follows:
\Beq
I_s (\varphi) & = & \frac{e}{2 h} \int^\infty_{-\infty} dE~{\rm{Tr}}\left[(\sigma_0 \tau_3)
\left(V_{ST}G^{+-}_{TS}-G^{+-}_{ST}V_{TS}\right)\right],  \label{eq15} 
\Eeq
where $\varphi = \phi_T - \phi_S$ is the superconducting phase difference across the junction and $G^{+-}_{ST/TS}$ are 
Keldysh-Green's functions, which at zero bias can be expressed as $G^{+-}_{ST/TS}(E) = n_{\rm F}(E) \,
[G^a_{ST/TS} (E) -G^r_{ST/TS}(E)]$. Here, $n_{\rm F}(E)$ is the Fermi-Dirac distribution at temperature $T$ and for chemical potential set to zero and $G^{a/r}_{ST/TS}(E)$ are advanced/retarded Green's function (GF) that describes 
the substrate-tip/tip-substrate ($ST/TS$) hybrid system. These GFs can be obtained via the Dyson equations $G^{r/a}_{ST} =
G^{r/a}_{SS}V_{ST}g^{r/a}_{TT}$ and $G^{r/a}_{TS} = g^{r/a}_{TT}V_{TS}G^{r/a}_{SS}$, where the GFs for the tip/substrate 
($TT/SS$) are given by $G^{r/a}_{TT/SS} = \left[{(g^{r/a}}_{TT/SS})^{-1} - \Sigma^{r/a}_{S/T} \right]^{-1}$ with 
self-energies $\Sigma^{r/a}_S = V_{ST} g^{r/a}_{TT} V_{TS}$ and $\Sigma^{r/a}_T = V_{TS} g^{r/a}_{SS} V_{ST}$. 
The tip/substrate hybridized GFs can be expressed in terms of the Pauli $\otimes$ Nambu matrices $\sigma_m\tau_n$ as 
$G^{r/a}_{jj}=\sum_{m,n}G^{r/a}_{jj;mn}\sigma_m\tau_n$ where $m,n=0,1,2$, and $3$. In general, Eq.~\eqref{eq15} has to be 
evaluated numerically. For a small tunneling parameter $t\ll \Gamma_{S/T}$ \cite{Villas2021}, away from the 0--$\pi$ transition 
lines, however, the current-phase relation can be obtained analytically as $I_s \sim \sin\varphi$ (see Appendix \ref{AppenA}). 
From a physical point of view, we find that the supercurrent is always carried by four phase-dependent 
bound states that result from the hybridization of the individual YSR states. Two of these states have negative energies and provide 
the dominant contribution to the current at low temperatures. The other two states have positive energies and start contributing to 
the supercurrent flow at higher temperatures.

\vspace{10 mm}

\section{Results and discussions} \label{sec-results}

In what follows, we illustrate the results for the supercurrent for different ranges of parameters and assume that both 
superconductors are identical with a gap equal to $\Delta_{S/T}(T)=\Delta_0\tanh(1.74{\sqrt{T_c/T-1}})$~\cite{gap}. At $T=0$ the superconductors 
($S/T$) are characterized by the BCS superconducting gap $\Delta_0 = 1.764k_{\rm B} T_c$ with the superconducting critical 
temperature $T_c$. For this purpose we define the critical current as $I_c = \left|I_s(\varphi_{\rm{max}})\right|$ where $|I_s|$ 
is the maximum at $\varphi=\varphi_{\rm{max}}$ such that $0\leq \varphi_{\rm{max}}\leq \pi$. On the other hand, we characterize the 0 
and $\pi$ phase of supercurrent as $I_s(\varphi_{\rm{max}})/I_c > 0$ and $I_s(\varphi_{\rm{max}})/I_c < 0$, respectively. Let us start 
by considering first the limit of low transparency (or weak coupling between the impurities). In Fig.~\ref{fig-scheme}(b) 
we present the results for the zero-temperature critical current as a function of the two exchange fields $J_S$ and $J_T$ for 
$t=\Delta_0$, $\theta=\pi/2$, $\Gamma_{S/T} = 100\Delta_0$ and $U_{S/T}=0$ \cite{note1}. Notice that there are up to four 0 and 
$\pi$ phases separated by boundaries that correspond to the YSR lying at zero energy (for $J_{S/T}/\Gamma_{S/T} = 1$ 
while $U_{S/T}=0$). In general, for low transparencies and sufficiently low temperatures, we find that the system exhibits 0 
and $\pi$ phases if $E_{YSR,S\sigma}E_{YSR,T\sigma}>0$ and $E_{YSR,S\sigma}E_{YSR,T\sigma}<0$, respectively. In this low 
transparency regime, the supercurrent is sinusoidal, i.e., $I_s \sim \sin\varphi$, almost for all ($J_S, J_T$) except near the 
0-$\pi$ transition lines [see Fig.~\ref{fig-scheme}(b)], where $I_s(\varphi)$ features higher harmonics, as we shall discuss below. 
At low temperatures and for low transparencies, the 0-$\pi$ features of the supercurrent, away from the transition lines, are fairly
independent of the value of $\theta$. At low temperatures and for low transparencies, if $U_{S/T}=0$, the critical current $I_c(J_S,J_T)$ 
exhibits avoided crossing of the $0$-$\pi$ transition lines at $J_{S}/\Gamma_{S}=J_{T}/\Gamma_{T}=1$ for $\theta\neq \pi/2$ or 
$3\pi/2$, e.g., $\theta=0$ and $\pi$ (see Appendix \ref{AppenB}). The avoided crossings become prominent for finite broadening $\gamma$.

\begin{figure}[t!]
\includegraphics[width=1\linewidth]{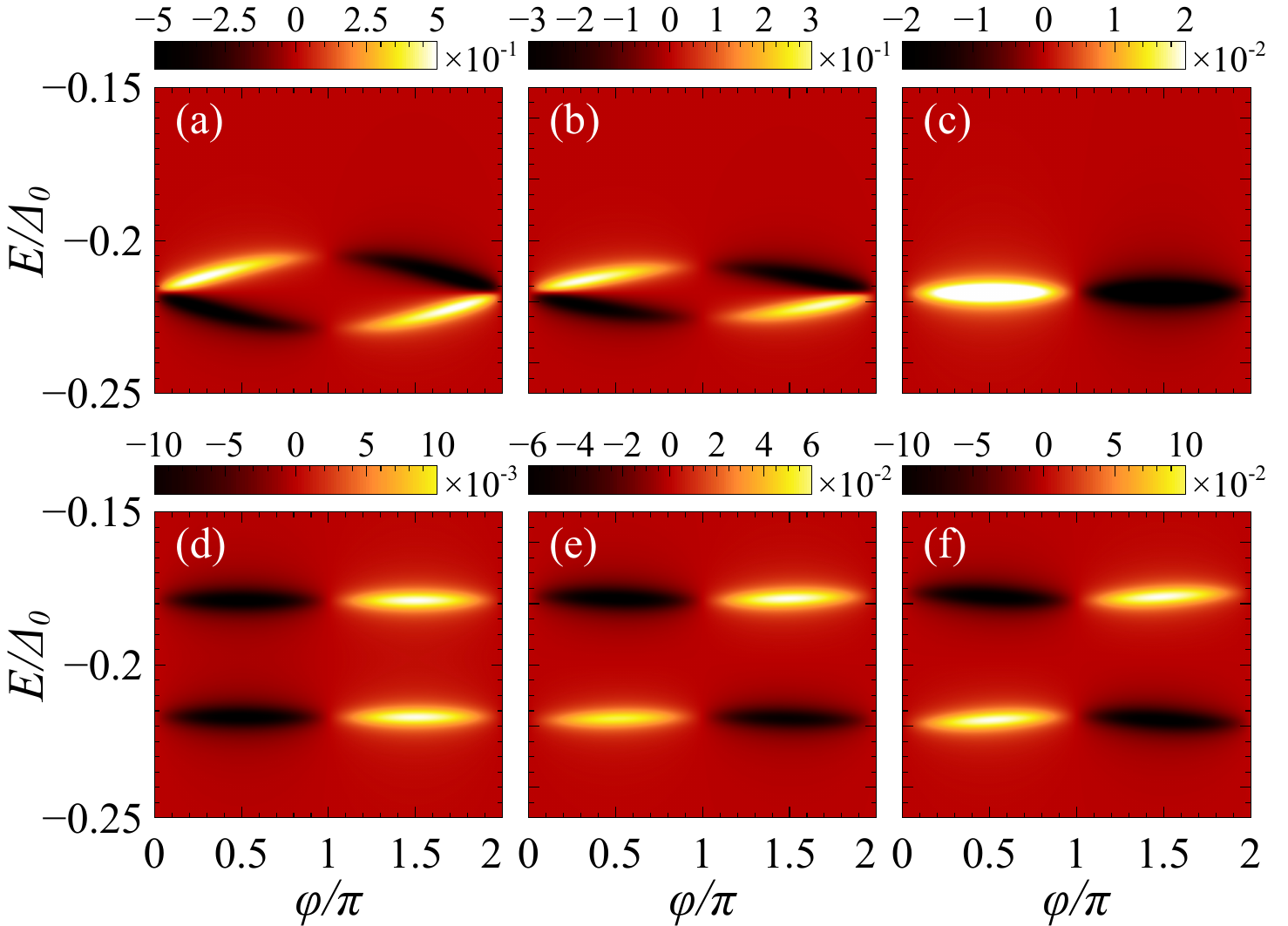}
\caption{\label{spectral}{  (a) The spectral current (in color maps),  $(\hbar/e)A(E,\varphi)$, for  $t=\Delta_0$, 
$\Gamma_{S/T}=100\Delta_0$, $U_{S/T}=0$, and $\gamma=0.003\Delta_0$ at $T=0$. Panels (a)-(c) are for $J_{S}/\Gamma_S=J_T/\Gamma_T=0.8$ 
[blue colored diamond shaped point in Fig.~\ref{fig-scheme}(b)] and panels (d)-(f) are for $J_S/\Gamma_S=1.2$ and $J_T/\Gamma_T=0.8$ [green colored circular shaped point in Fig.~\ref{fig-scheme}(b)]. The exchange fields orientation angle $\theta$ is set to $0$ (a), (d); 
$\pi/2$ (b), (e); and $\pi$ (c), (f). Panel (c) exhibits two degenerate bound states due to the symmetry of the two bare systems.}}
\end{figure}

To illustrate the $\theta$ and $T$ dependencies of critical current for low transparencies we plot $I_c(\theta, T)$ for
$E_{YSR,S\sigma}E_{YSR,T\sigma}>0$ and $E_{YSR,S\sigma}E_{YSR,T\sigma}<0$ cases in Figs.~\ref{fig-scheme}(c) and~\ref{fig-scheme}(d), 
respectively. We see that $I_c$ increases at low $T$ from $\theta=0$ to $\pi$ in Fig.~\ref{fig-scheme}(c) and vice versa in 
Fig.~\ref{fig-scheme}(d). Figure~\ref{fig-scheme}(c) also illustrates that $0$-$\pi$ transitions (at a fixed $\theta$) can be 
induced by increasing the temperature for $0 \leq \theta < \pi/2$ and $3\pi/2 < \theta \leq 2\pi$ if $E_{YSR,S\sigma}
E_{YSR,T\sigma}>0$. On the other hand, a $\pi$-$0$ transition appears in Fig.~\ref{fig-scheme}(d) with increasing $T$ for 
$\pi/2 < \theta < 3\pi/2$ if $E_{YSR,S\sigma}E_{YSR,T\sigma} < 0$. We notice that the critical angle $\theta=\theta_c$, at 
which a 0-$\pi$ transition occurs, depends on $T$. For low transparencies we do not find temperature-induced 0-$\pi$ 
transitions for $\theta=\pi/2$ or $3\pi/2$, see Figs.~\ref{fig-scheme}(c) and~\ref{fig-scheme}(d). Physically, upon increasing 
the temperature, the positive-energy bound states start to contribute to the supercurrent. It can be shown that the total current 
carried by the two negative-energy states is in the reverse direction of the total current carried by the two positive-energy 
states. The 0-$\pi$ transitions in Figs.~\ref{fig-scheme}(c) and~\ref{fig-scheme}(d) are due to the enhanced population of the 
higher energy states with increasing temperature. In Figs.~\ref{fig-scheme}(c) and \ref{fig-scheme}(d) we notice that for higher
temperatures there are also $\theta$-dependent 0-$\pi$ transitions.

In what follows, we report spectral current, $A(E,\varphi)$, features at low $T$ and for low transparencies while $J_{S/T}\neq 0$ 
are away from the transition lines. In general, the supercurrent can be expressed as $I_s(\varphi)=\int dE~A(E,\varphi)$. At 
low temperatures, only the two negative-energy bound states of the hybrid system contribute to the supercurrent [see 
Figs.~\ref{spectral}(a)-\ref{spectral}(f)]. We find in Fig.~\ref{spectral}(a) for $E_{YSR,S\sigma}E_{YSR,T\sigma}>0$ and $\theta=0$ 
that the two negative-energy states carry the current in opposite directions. In this case, it can be shown that they together 
result in a $0$ phase, see, e.g., Fig.~\ref{fig-scheme}(c) for $\theta=0$ at low temperatures. For $E_{YSR,S\sigma}E_{YSR,T\sigma}
>0$ the magnitude of the relative contribution to the supercurrent monotonically increases as $\theta$ changes from $0$ to $\pi$ 
[see Figs.~\ref{spectral}(a)-\ref{spectral}(c)]. In fact, in this case for $\theta=\pi$ the two negative energy states 
carry the current in the same direction [see Fig.~\ref{spectral}(c)]. Consequently, for $E_{YSR,S\uparrow}E_{YSR,T\uparrow}>0$ at 
low $T$ the critical current monotonically increases from $\theta=0$ to $\pi$ featuring $0$ phases [see Fig.~\ref{fig-scheme}(c) 
at low temperatures]. On the other hand, for $E_{YSR,S\sigma}E_{YSR,T\sigma}<0$ and $\theta=\pi$ we find in Fig.~\ref{spectral}(f) 
that the two negative energy states carry the current in opposite directions. In this case, it can be shown that they together 
result in a $\pi$-phase, see, e.g., Fig.~\ref{fig-scheme}(d) for $\theta=\pi$ at low temperatures. In this case, the magnitude of 
the relative current increases as $\theta$ changes from $\pi$ to $0$ [see Fig.~\ref{spectral}(d)-\ref{spectral}(f)]. In 
fact, in this case for $\theta=0$ the two states carry current in the same direction [see Fig.~\ref{spectral}(c)]. Consequently, 
for $E_{YSR,S\uparrow}E_{YSR,T\uparrow}<0$ at low $T$ the critical current monotonically increases from $\theta=\pi$ to $0$ 
featuring $\pi$-phases [see Fig.~\ref{fig-scheme}(d) at low temperatures].
\begin{figure*}[t!]
\includegraphics[width=0.95\linewidth]{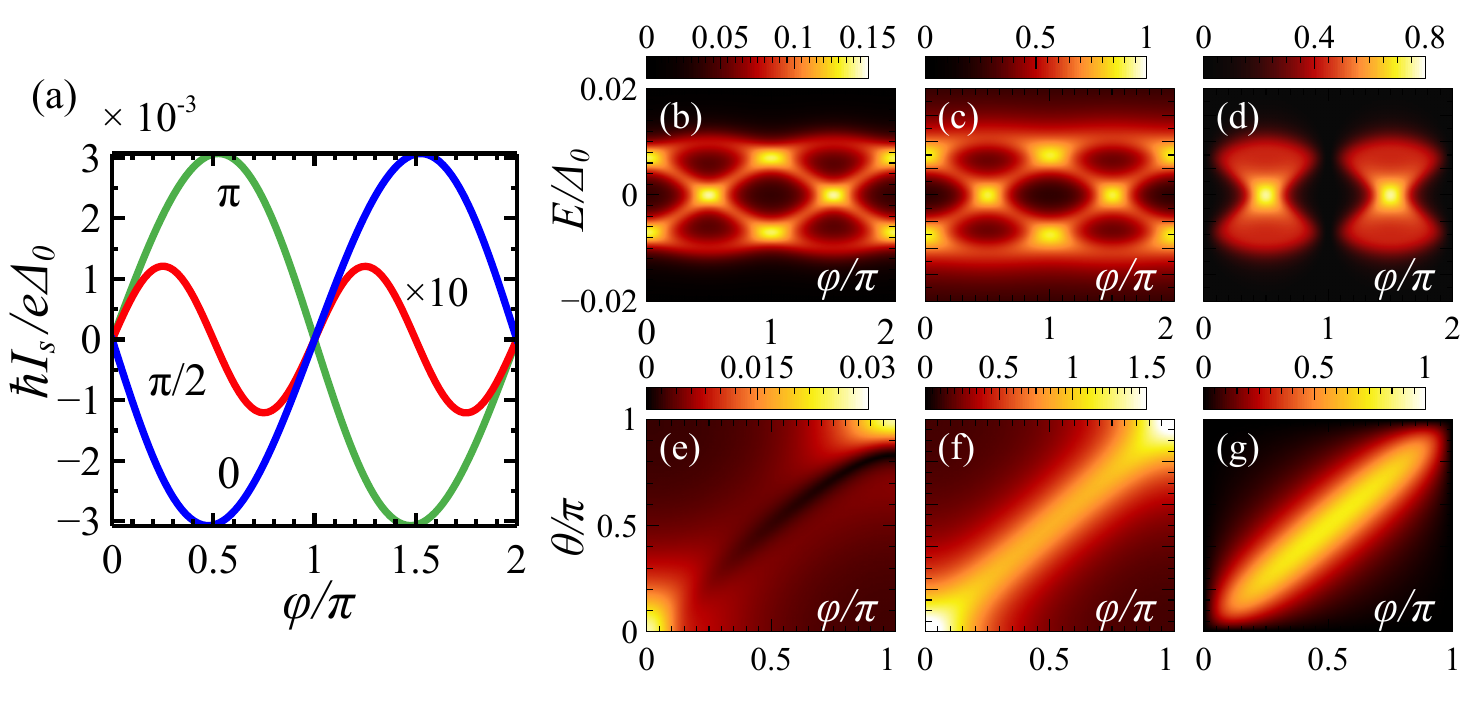}
\caption{\label{triplet}{ (a) Current-phase relation $I_s(\varphi)$ for different $\theta$ (indicated in the plot), 
$t=\Delta_0$, $J_{S}/\Gamma_{S}=J_{T}/\Gamma_{T}=1$, $\Gamma_{S/T}=100\Delta_0$, and $U_{S/T}=0$ at $T=0$. (b) Local density of states 
(LDOS), (c) mixed-triplet pairing amplitude, and (d) pure-triplet pairing amplitude for the parameters of the red curve in panel (a). 
The zero-energy modes of (e) singlet pairing, (f) mixed triplet pairing, and (g) pure triplet pairing amplitudes for $t=\Delta_0$,
$J_{S}/\Gamma_{S}=J_{T}/\Gamma_{T}=1$, $\Gamma_{S/T}=100\Delta_0$, and $U_{S/T}=0$ at $T=0$. In all panels $\gamma=0.003\Delta_0$. 
The quantities with colormaps in panels (b)-(g) are expressed in the units of $\Delta^{-1}_0$. }}
\end{figure*} 
\begin{figure}[b!]
\includegraphics[width=1\linewidth]{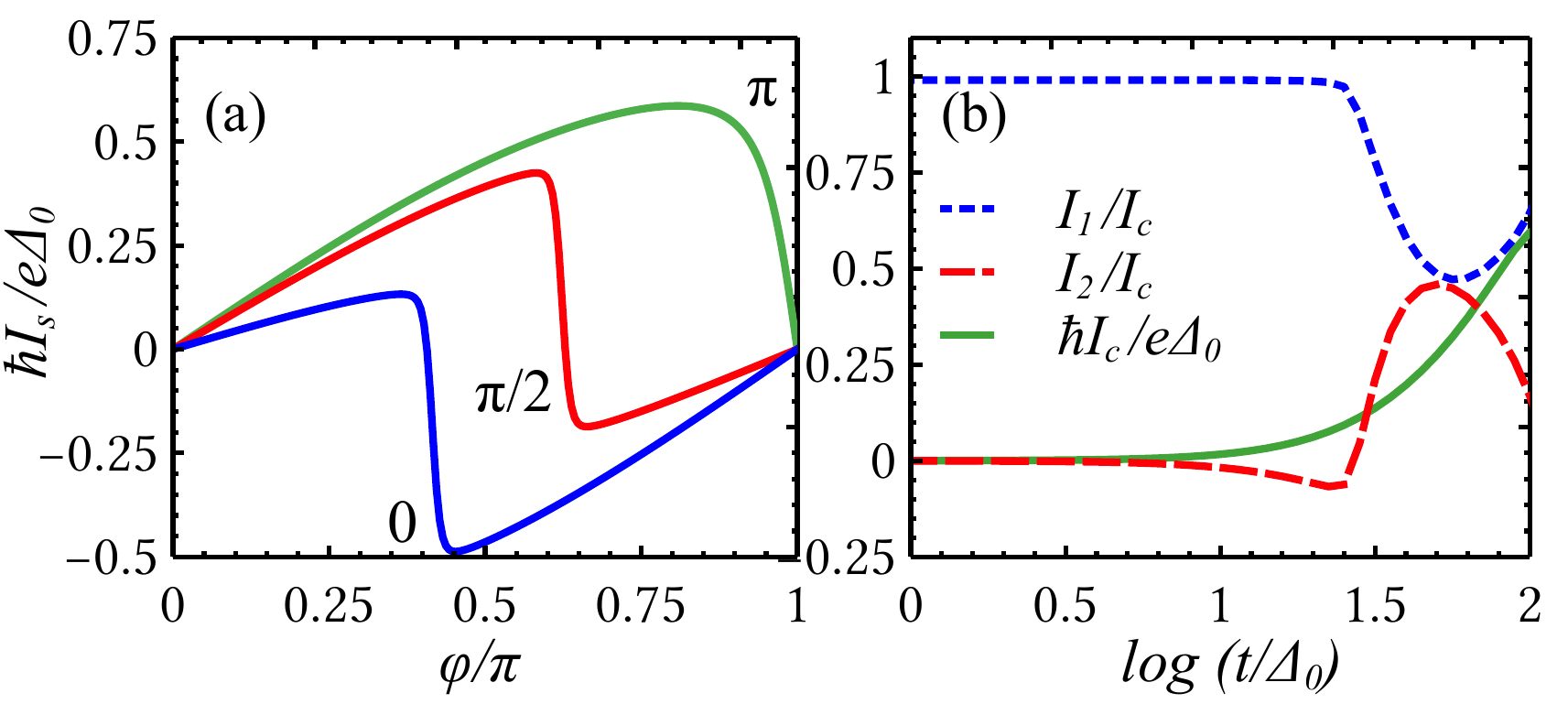}
\caption{  (a) Zero-temperature current-phase relation for different $\theta$ (indicated in the plot), 
$t=70\Delta_0$, $\Gamma_{S/T}=100\Delta_0$, $U_{S/T}=0$, and $\gamma=0.003\Delta_0$, and $J_{S}/\Gamma_{S}= J_{T}/\Gamma_{T}=0.8$. 
(b) The first (dotted blue) and second (dashed red) harmonics of the current-phase relation for $\theta=\pi/2$ as a function of the hopping 
$t$, as well as the critical current $I_c$ (solid green curve). The rest of the parameters are the same as in panel (a).} 
\label{fig-harmonics}
\end{figure} 

Let us remind that usually $I_s \sim \sin\varphi$ for low transparencies, except near the 0-$\pi$ transition lines in the $I_c(J_S,J_T)$
diagram [see, e.g., Fig.~\ref{fig-scheme}(b)]. We now show that at low temperatures higher harmonics in $I_s(\varphi)$ can appear near 
the 0-$\pi$ transition lines. Figure~\ref{triplet}(a) shows such higher harmonics at the intersection of the transition lines in 
Fig.~\ref{fig-scheme}(b), i.e., for $\theta=\pi/2$ and for $E_{{\rm YSR},{S}\sigma}=E_{{\rm YSR},{T}\sigma} = 0$. The appearance of 
higher harmonics indicates that there can be supercurrent due to pure (equal spins) triplet pairing. In this case,
and despite the fact that $t\ll\Gamma_{S/T}$, we find that the higher powers of tunneling parameters, beyond $t^2$, 
significantly contribute to supercurrent. In particular, with the set of 
parameters in Fig.~\ref{triplet}(a), for which $E_{{\rm YSR},{S}\sigma}=E_{{\rm YSR},{T}\sigma} = 0$ and $\theta=\pi/2$, we observe 
that $I_s(\varphi)$ features a pure second harmonic.  At this point, we always find that $I_s(\varphi)$ features a pure second 
harmonic at low temperatures and for arbitrarily low transparencies. Higher harmonic features for $E_{{\rm YSR},{S}\sigma} = 
E_{{\rm YSR},{T}\sigma} = 0$ diminish as $\theta$ changes from orthogonal to parallel/antiparallel orientations. 
Figure~\ref{triplet}(a) also shows that $I_s(\varphi)$ for $\theta=0$ and $\pi$ when $E_{{\rm YSR},{S}\sigma} = 
E_{{\rm YSR},{T}\sigma} = 0$, which exhibits features of $\pi$ and 0 phases for $\theta=0$ and $\pi$, respectively. These suggest 
that for $\theta=0$ two diagonal $\pi$-phase blocks, associated with $E_{YSR,S\sigma}E_{YSR,T\sigma}<0$, and for $\theta=\pi$ two 
diagonal $0$-phase blocks, associated with $E_{YSR,S\sigma}E_{YSR,T\sigma}>0$, in $I_c(J_S,J_T)$ diagram would be continuously joined 
at $J_{S}/\Gamma_{S}=J_{T}/\Gamma_{T}=1$ while $U_{S/T}=0$. Consequently, $I_c(J_S,J_T)$ would exhibit avoided crossing of $0$-$\pi$ 
transition lines for $\theta=0$ and $\pi$ (see Appendix \ref{AppenB}). 

Comparing  Figs.~\ref{triplet}(b) and ~\ref{triplet}(d) for $\theta=\pi/2$ and $E_{{\rm YSR}, S \sigma}=E_{{\rm YSR}, T \sigma}=0$, 
we observe that the bound state crossing in the density of states ${\rm{LDOS}} = {\rm{Im}}({\rm{Tr}}[G^a_{SS}+G^a_{TT}])/16\pi$ 
is responsible for the onset of pure triplet pairing. The pure triplet pairing amplitude for the substrate 
in terms of the hybridized GF's components is obtained as $T_{S;\uparrow\uparrow} = |(G^a_{SS;11}-G^a_{SS;22})-
i(G^a_{SS;12}+G^a_{SS;21})|$. The appearance of pure triplet superconductivity makes $I_s(\varphi)$ non-sinusoidal. 
Figures~\ref{triplet}(b),~\ref{triplet}(c), and~\ref{triplet}(d) show the LDOS, the substrate's mixed (different spins) triplet 
pairing amplitude, and the substrate's pure triplet pairing amplitude, respectively, for $\theta=\pi/2$. The substrate's mixed triplet 
pairing amplitude is obtained in terms of the hybridized GF's components as $T_{S;\uparrow\downarrow} = |(G^a_{SS;31}-iG^a_{SS;32})|$. 
In Figs.~\ref{triplet}(e)-\ref{triplet}(g) we display $\theta$ and $\varphi$ dependencies on zero-energy modes of substrate's singlet 
pairing, mixed triplet pairing, and pure triplet pairing amplitudes for relatively weak coupling between the impurities, $t=\Delta_0$, 
setting $J_{S}/\Gamma_{S} = J_{T}/\Gamma_{T}=1$, and $U_{S/T}=0$. The substrate's singlet pairing amplitude in terms of the hybridized 
GF's components is obtained as $S_{S;\uparrow\downarrow} = |(G^a_{SS;01}-iG^a_{SS;02})|$. We see maximum pure triplet pairing amplitude 
for $\theta=\pi/2$ and $\varphi=\pi/2$. This features a pure second harmonic in $I_S(\varphi)$ for orthogonal orientation when 
$E_{{\rm YSR}, S\sigma}=E_{{\rm YSR}, T\sigma}=0$. Higher harmonics and pure triplet features become important for arbitrary $J_{S/T}$ 
and for all $\theta$ when the tunneling parameter $t$ is large.

Finally, we illustrate the contribution of the higher harmonics of $I_s(\varphi)$ in Fig.~\ref{fig-harmonics}(a) for a higher 
transparency $t = 70\Delta_0$ and various $\theta$ values. The results of Fig.~\ref{fig-harmonics}(a) were obtained for 
$J_{S}/\Gamma_{S}= J_{T}/\Gamma_{T}=0.8$, which is set away from the 0-$\pi$ transition lines in $I_c(J_S,J_T)$. We observe that 
higher harmonics are more prominent for $\theta=\pi/2$. In general, supercurrent can be expressed in terms of higher harmonics as
$I_s(\varphi)=\sum_nI_n\sin (n\varphi)$ with positive integers $n$. Figure~\ref{fig-harmonics}(b) shows how the first and 
second harmonics vary with increasing transparencies $t$. We clearly see that the second harmonic first 
enhances with increasing $t$ from low transparencies. For moderate transparencies, the second harmonic plays a significant role 
in $I_s(\varphi)$. In this case, pure triplet also has a significant contribution to the supercurrent. 
\section{Conclusions} \label{sec-conclusions}

Inspired by recent STM-based experiments, we have presented a theoretical investigation of
the DC Josephson effect between two magnetic impurities coupled to superconductors such that they host YSR states.
We have shown that the supercurrent in this system exhibits a very rich phenomenology. For instance, for weak coupling 
between the magnetic impurities, the current-phase relation is sinusoidal, but exhibits various types of $0$-$\pi$ transitions
due to changes in the YSR energies, temperature, and orientation of the impurity magnetic moments. Upon increasing the coupling,
the current-phase relation becomes nonsinusoidal due to the appearance of (pure) triplet superconductivity. Our results can be 
tested experimentally in the context of STM and magnetic impurities on superconducting surfaces \cite{Huang2020a,Huang2021}.


\acknowledgements

S.C., D.N., and W.B.\ acknowledge support from the EU Horizon 2020 research and innovation program under Grant Agreement No.\ 
964398 (SUPERGATE) and from the Deutsche Forschungsgemeinschaft (DFG; German Research Foundation) via SFB 1432 (Project No. 425217212).
R.S.S.\ acknowledges funding from the Spanish CM "Talento Program" (Project No.\ 2022-T1/IND-24070) and the European Union Horizon 
2020 research and innovation program under the Marie Sklodowska-Curie Grant Agreement No.\ 10103324. J.C.C.\ thanks the Spanish Ministry 
of Science and Innovation (Grant PID2020-114880GB-I00) for financial support and the Deutsche Forschungsgemeinschaft (DFG; German Research Foundation) and SFB 1432 for sponsoring his stay at the 
University of Konstanz as a Mercator Fellow. 

\vspace{-4mm}

\appendix

\section{Supercurrent in the tunnel regime} \label{AppenA}

In the appendix we elaborate on the analysis of the supercurrent in the tunneling regime and, in particular, we provide some analytical
insight into the current-phase relation and the critical current in this regime.

The tip/substrate ($TT/SS$) bare GFs can be decomposed in the spin $\otimes$ Nambu space as follows
\Beq
 {g}^a_{jj} &=&  \sum_{l=0,3}  \sum_{m=0,3} {g}^a_{jj;lm}~\sigma_l\tau_m  +  \sum_{l=0,3}  \sum_{m=+,-} {g}^a_{jj;lm}~\sigma_l\tau_m ,~
\label{eq12a}
\Eeq
with $\tau_{\pm}=(\tau_1 \pm i\tau_2)/2$ and ${g}^a_{jj;l\pm}={f}^a_{jj;l +}e^{\pm i\phi_j}$. The first and second terms in 
Eq.~\eqref{eq12a} represent the normal and anomalous components of the GF, respectively. In the anomalous part $f^a_{jj;0+}$ 
and $f^a_{jj;3+}$ are for the singlet and mixed-triplet contributions, respectively. Using Eq.~\eqref{eq15} we can now express 
the supercurrent for low transparencies as
\Beq
I_s(\varphi) &=& 8t^2 \frac{e}{h} \sin\varphi \int^\infty_{-\infty} dE~n_F(E)~ {\rm{Im}}\left[f^a_{SS;0+}f^a_{TT;0+}  \right. \nonumber \\
&& \left. + f^a_{SS;3+}f^a_{TT;3+}\cos\theta\right].   \label{eq18}
\Eeq
We see that the current-phase relation is sinusoidal $I_s\sim\sin\varphi$ in this limit. Furthermore, we can consider 
$J_j,\Gamma_j\gg|E_{YSR,j\uparrow}|$ and approximate the denominator of the bare Green's functions as follows
\begin{equation} \label{eq-D}
D_{j\sigma}(E) \approx -{\rm{sign}}(J_{j\sigma})\psi_j ~(E- {\rm{sign}}(J_{j\sigma}) E_{YSR,j\uparrow}), \\
\end{equation}
where
\begin{equation}
\psi_j = \frac{1}{2\Gamma_jJ_j} \left[(J^2_j-U_j^2-\Gamma^2_j)^2 +4\Gamma^2_j J^2_j \right]. 
\end{equation}
With this approximation, we can obtain the approximate analytical expression of supercurrent for low transparencies as
\Beq
I_s(\varphi) &=& \left(\frac{8et^2\Gamma_S\Gamma_T\Delta_S\Delta_T}{h\psi_S\psi_TE^+_{YSR}E^-_{YSR}}\right) 
\left[E^+_{YSR} ~\rho^-_{YSR} \cos^2(\theta/2) \right. \nonumber \\
&& \left. + E^-_{YSR}~(1-\rho^+_{YSR})~\sin^2(\theta/2)\right] \sin\varphi , \label{eq21}
\Eeq
where $E^\pm_{YSR} = E_{YSR,S\uparrow}\pm E_{YSR,T\uparrow}$ and $\rho^{\pm}_{YSR} = n_{\rm F}(E_{YSR,S\uparrow}) 
\pm n_{\rm F}(E_{YSR,T\uparrow})$. This formula qualitatively captures $T$- and $\theta$-dependencies of the supercurrent. 
At low $T$, this analytical formula gives a vanishing supercurrent for $\theta=0$ and $\theta=\pi$ when $E_{YSR,S\uparrow}
E_{YSR,T\uparrow}>0$ and $E_{YSR,S\uparrow}E_{YSR,T\uparrow}<0$, respectively. This is not correct due to the approximation 
of $D_{j\sigma}$. The situation is more complicated if $E_{YSR,S\sigma} E_{YSR,T\sigma}=0$.

\begin{figure}[t!]
\includegraphics[width=1\linewidth]{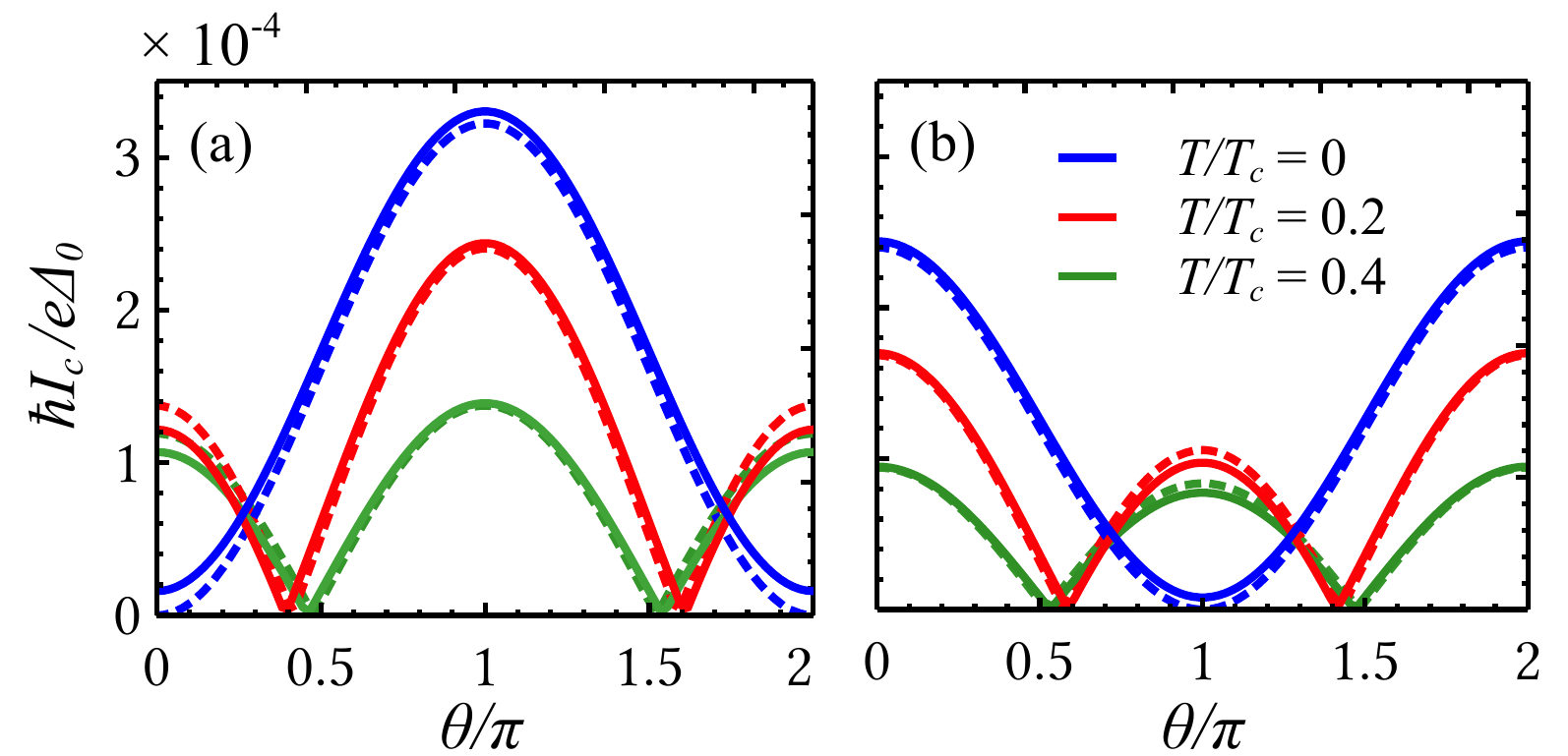}
\caption{\label{compare}{ $I_c(\theta)$ for (a) $J_{S/T}/\Gamma_{S/T}=0.8$, and (b) $J_{S}/\Gamma_{S}=0.8$ and $J_{T}/\Gamma_{T}=1.2$. 
For all panels $\Delta_S=\Delta_T$, $\Gamma_{S/T}=100\Delta_0$, $t=\Delta_0$, $U_{S/T}=0$, and $\gamma=0.003\Delta_0$. The solid lines 
were obtained with Eq.~\eqref{eq15} (the exact numerical approach) and the dashed lines correspond to the results obtained with the 
analytical formula of Eq.~\eqref{eq21}. }}
\label{fig1}
\end{figure}

\begin{figure}[b!]
\includegraphics[width=1\linewidth]{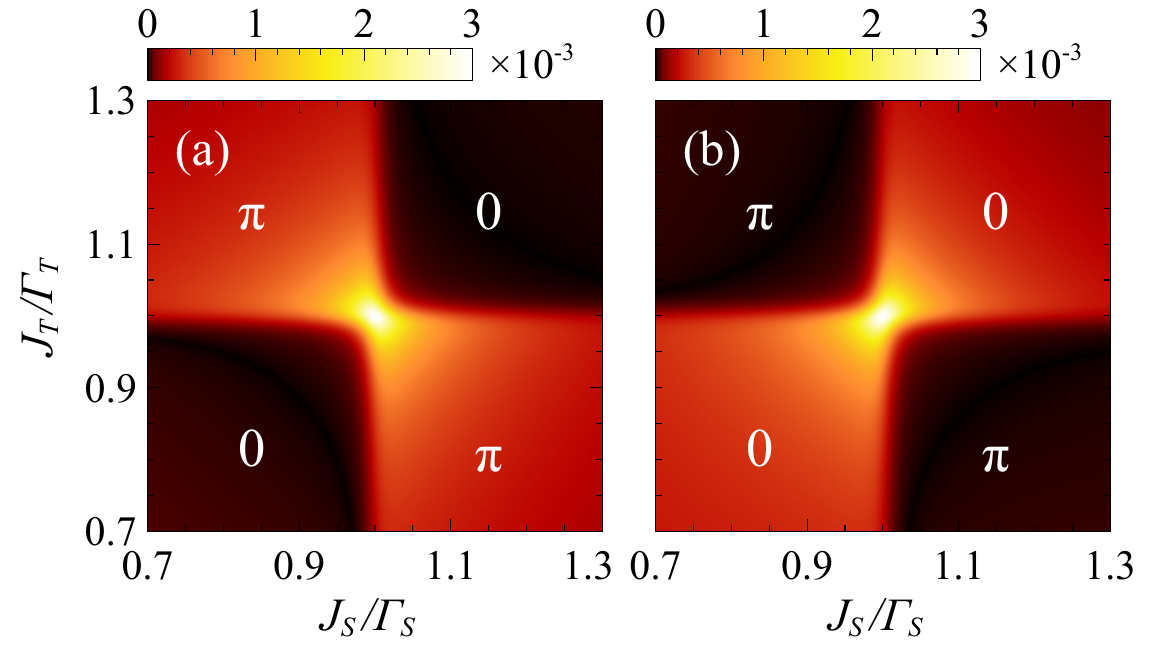}
\caption{\label{ICJJ}{The zero temperature critical current as a function of $J_T$ and $J_S$ for $\theta$ values (a) $0$ 
and (b) $\pi$, respectively. For all panels $\Delta_{S}=\Delta_T$, $\Gamma_{S/T}=100\Delta_0$, $t=\Delta_0$,
$U_{S/T}=0$, $\gamma=0.003\Delta_0$ and $T=0$. The critical currents with colormaps in panels (a) and (b) are expressed in the units of 
$e\Delta_0/\hbar$.}}
\end{figure} 


In Fig.~\ref{compare} we present the results for the critical current $I_c=|I_s(\varphi_{\rm{max}})|$ computed with Eq.~\eqref{eq15} 
(solid lines) and Eq.~\eqref{eq21} (dashed lines) as a function of the relative orientation $\theta$ of the two magnetic impurities at 
various temperatures. We note that $|I_s(\varphi)|$ is maximum at $\varphi=\varphi_{\rm{max}}$. For the set of parameters in 
Fig.~\ref{compare}(a) we find a $0 \rightarrow \pi$ phase shift for $0 \leq \theta < \pi/2$ and $3\pi/2 < \theta \leq 2\pi$ with 
increasing temperatures for $E_{YSR,S\sigma}E_{YSR,T\sigma}>0$. For the set of parameters in Fig.~\ref{compare}(b) we find a 
$\pi \rightarrow 0$ phase shift for $\pi/2 <\theta < \pi$ with increasing temperature for $E_{YSR,S\sigma}E_{YSR,T\sigma}>0$. 
The analytical formula in Eq.~\eqref{eq21} works well for $J_{S/T}$ values away from the 0-$\pi$ transition lines in the$I_c(J_S,J_T)$ diagram. 


\section{Critical current for nonorthogonal orientations} \label{AppenB}

We now consider the behavior of the critical current as a function of the exchange energies, $I_c(J_T,J_S)$, for nonorthogonal 
configurations ($\theta \neq \pi/2$) at zero temperature and for low coupling $t\ll\Gamma_{S/T}$. In Fig.~\ref{ICJJ}(a) and (b) we present 
$I_c(J_S,J_T)$ for $t=\Delta_0$, $\Gamma_{S/T} = 100\Delta_0$, $U_{S/T}=0$, and for $\theta=0$ and $\pi$, respectively. Usually, for 
low transparencies and low enough temperatures, we find that the system exhibits $0$ and $\pi$ phases for all $\theta$ if
$E_{YSR,S\sigma}E_{YSR,T\sigma}>0$ and $E_{YSR,S\sigma}E_{YSR,T\sigma}<0$, respectively. However, very close to the $0$-$\pi$ 
transition lines in $I_c(J_T,J_S)$ the supercurrent features vary with $\theta$ [compare Figs.~\ref{fig-scheme}(b), \ref{ICJJ}(a), 
and \ref{ICJJ}(b)]. We see that $I_c(J_T,J_S)$ shows distinct avoided crossing of $0$-$\pi$ transition lines for $\theta=0$ and 
$\pi$ in Figs.~\ref{ICJJ}(a) and \ref{ICJJ}(b), respectively. This avoided crossing becomes more prominent for finite spectral 
broadening, $\gamma$. In contrast, no avoided crossing happens between the $0$-$\pi$ transition lines for $\theta=\pi/2$ 
[see Fig.~\ref{fig-scheme}(b)]. The magnitude of the current strongly depends on the spin of the hybridized YSR states. 
For $\theta=0$ transport through parallel oriented negative energy, YSR states are suppressed due to the spin polarization: electrons 
with opposite spins have to virtually occupy states with very high energy. For $\theta=0$, the two negative energy YSR states have 
parallel spins in the 0 phase, where $E_{YSR,S\sigma}E_{YSR,T\sigma}>0$. In contrast, the two negative energy YSR states have opposite 
spin in the $\pi$ phase, where $E_{YSR,S\sigma}E_{YSR,T\sigma}<0$, which makes the critical current in the $\pi$ phase larger than 
in the $0$ one, see Fig.~\ref{ICJJ}(a). The situation is reversed for $\theta=\pi$, as illustrated in Fig.~\ref{ICJJ}(a). In the 
intermediate situation ($\theta=\pi/2$) critical currents for 0 and $\pi$ phases are identical.

\end{document}